\documentclass [fleqn,12pt]{article}
\usepackage{amsmath,amssymb,latexsym,cite}
\usepackage[english]{babel}
\usepackage{graphicx}
\usepackage{epstopdf}
\usepackage{caption}
\usepackage{subcaption}
\usepackage[T2A]{fontenc}
\usepackage[utf8]{inputenc}
\usepackage{amsmath,amssymb,cmap,mathrsfs,indentfirst,graphicx,cite,geometry,titlesec,setspace,longtable,hyperref}

\title{Development of dark disk model of positron anomaly origin}
\author{K.M. Belotsky$^a$, A.A. Kirillov$^{a,b}$ and M.L. Solovyov$^a$}
                \date{\small\it
	\begin{description}
		\item[$^a$] 
		National Research Nuclear University MEPhI 
		(Moscow Engineering Physics Institute), Moscow, Russia;\\
        \item[$^b$] Yaroslavl State P.~G.~Demidov University, Yaroslavl, Russia
	\end{description}    
}

\begin{document}
 
\maketitle

    \begin{abstract}
		Dark disk model could be a
		remedy for dark matter (DM) explanation of positron anomaly (PA) in cosmic rays (CR). The main difficulty in PA explanation relates to cosmic gamma-radiation which is inevitably produced in DM annihilation or decay leading to tension with respective observation data. Introduction of ``active'' (producing CR) DM component concentrating in galactic disk alleviates this tension. 
        Earlier we considered two-lepton modes, with branching ratios being chosen to fit in the best way all the observation data. Here we considered, in framework of the same dark disk model, two cases: two-body final state annihilation and four-body one, and in each case a quark mode is added to the leptonic ones. It is shown that 4-body mode case is a little better than 2-body one from viewpoint of quality of observation data description at the fixed all other parameters (of CR propagation, background, disk height). The values of DM particle mass around 350 GeV and 500 GeV are more favourable for 2- and 4-body modes respectively. Higher values would improve description of data on positrons only but accounting for data on gamma-radiation prevents it because 
        of unwanted more abundant high-energy gamma production. Inclusion of the quark modes improves a little fitting data in both 4- and 2-body mode cases, contrary to naive expectations. In fact, quark mode has a bigger gammas yield than that of most gamma-productive leptonic mode~--- tau, but they are softer due to bigger final state hadron multiplicity. 
        
    \end{abstract}
    
    \section{Introduction}
There is a plethora of hypotheses aimed to solution of dark matter (DM) problem. Many of them are being connected in investigation to other problems of astrophysics or cosmology, which solution could be reached by attributing to DM particles some specific properties. Cosmic ray (CR) puzzles are popular subject of such investigations.
It began with the first WIMP candidate \cite{bib:ZeldKhlop}, which was used for explanation of CR data \cite{bib:KonoplKhlop}, and now many other DM candidates are probed for theses purposes.
 
Positron anomaly in CR discovered by experiment PAMELA \cite{Pamela} and confirmed with high accuracy by AMS-2 \cite{AMS} is a shining example here \cite{Review}. It gave rise to a bulk of attempts of the anomaly explanation with different DM candidates. Now the rate of these attempts subsided noticeably 
because of the emerged constraints coming from cosmic gamma-radiation and cosmic microwave background (CMB). CMB constraint \cite{planck} is applied for annihilating DM. One can try to avoid it, making a tuning with narrow resonance in DM annihilation \cite{reson-1, reson-2}, or $p$-wave annihilation \cite{bib:Diamanti}, or adjusting two 
dark species with one decaying (after recombination) into another which annihilates in Galaxy \cite{decann}, or maybe something other. But all such attempts along with decaying DM scenario face difficulty in compatibility with data on gamma-radiation. They are Fermi/LAT data on Isotropic Gamma-Radiation Background (IGRB) \cite{fermi-igrb}, gamma-radiation from Galactic Center (GC) \cite{fermi-GC}, from dwarf galaxies \cite{fermi-dwarf}. Other data (e.g. for other galactic coordinates) are either irrelevant or give weaker constraint.

To avoid aforementioned constraints we suggested 
\cite{dd-1, dd-1_1, dd-2, dd-3, dd-4, dd-5} that dark matter may contain a small component which concentrates in galactic disk and produces cosmic rays due to annihilation or decay. 
In this case high latitude gamma-ray radiation from DM is suppressed due to a decrease of volume where from DM induced gamma-rays come.
It is worth noting that dark disk model was suggested and discussed from independent considerations of DM dynamics \cite{dd-6, dd-7, dd-8, dd-9}, though it is as a rule assumed to be more thin there. 

Earlier for dark disc model we considered leptonic modes calculating contributions to cosmic positrons, IGRB, gamma from GC, and also from galaxy Andromeda. In this work we add quark mode, include 4-body modes and, in addition to data on cosmic positrons, IGRB and gamma from GC, take into account data on cosmic antiprotons. The aim is to inquire how quality of data description ($\chi^2$) changes (spoils or improves) if quark mode, 4-body final states and antiproton data are involved in analysis.

To finish introduction one notes that there are alternative explanations of positron anomaly with the help of pulsars (e.g. \cite{puls}, or on the last situation \cite{bib:Abeysekara, bib:HooperLinden}) or super-Novae (e.g. \cite{ss}). To distinguish them from DM explanation is certainly important task \cite{calet} and it is in progress. Also specifics of possible anisotropy of CR propagation in galactic magnetic fields \cite{bib:Giacinti1, bib:Giacinti2} may effectively shorten CR travelling length. So CR puzzles could be explained \cite{bib:Kachelriess, bib:Savchenko, bib:Kachelrie} due to not (only) DM but with nearest SN \cite{bib:Wallner, bib:Fimiani}.

\section{CR analysis}
Our analysis is based on methods elaborated in our previous works \cite{dd-1, dd-2, dd-3, dd-4, dd-5}. 

One shortly describes them. Injection spectra from DM annihilation are simulated with the code Pythia \cite{pythia}, effects of propagation are calculated by GALPROP \cite{galprop}. CR propagation parameters are taken from \cite{propar}. Background fluxes for $e^{\pm}$ component are taken from \cite{backpos}. Estimation of the background for high-energy gamma-radiation from GC is model dependent and we did not take into account it. Though one notes that accounting for background contribution here spoils $\chi^2$. For analysis here we included the last (in energy) 6 datapoints for total gamma-flux \cite{fermi-GC}.

Antiprotons is also involved in analysis here. Observation data and background for them were taken from \cite{AMScite}. We included in analysis $11$ datapoints spread between $20$ and $300$ GeV.

DM density distribution was set in the form of $\propto \exp(-r/r_c)\exp(-|z|/z_c)$ like in \cite{dd-5, dd-6}, where the disk height $z_c=0.4$ kpc taken from the previous analysis.

The main formula is $\chi^2$ to be minimized
	\begin{equation}\label{chiSq}
	\chi^2 = \chi_\text{pos}^2+\chi_\text{IGRB}^2+\chi_\text{GC}^2+\chi_\text{antip}^2,
	\end{equation}
where each term relates to respective data (on positron fraction, IGRB, gamma from GC and antiprotons). For positrons we have
\begin{equation}
\label{chipos}
\chi_\text{pos}^2=\sum_{i=1}^{N_\text{pos}}\left( \dfrac{ F(E_i)-F_i^\text{exp} }{ \sigma_i} \right)^2,
\end{equation}
while for the rest
\begin{equation}
\label{chispes}
\chi_\text{species}^2=\sum_{j=1}^{N_\text{species}} \eta\left( \Phi(E_j)-  \Phi_j^\text{exp} \right) \left( \dfrac{  \Phi(E_j)-  \Phi_j^\text{exp} }{ \sigma_j } \right)^2.
\end{equation}
Here $F$ and $\Phi$ are the fluxes, $\sigma_i$ is the error, $\eta$ is the step function, the sum is over experimental datapoints, ${N_\text{pos}}$ and $N_\text{species}={N_\text{IGRB,GC,antip}}$ are the numbers of respective datapoints involved in analysis.
The main difference between \eqref{chipos} and \eqref{chispes} is existence of the step function in \eqref{chispes}, what implies requirement not to exceed experimental points for all data except for positrons, which should be fitted. In case of data on antiprotons, one could require both not exceeding and fitting data. But the latter makes little sense taking into account existing large uncertainties in background (secondary) antiproton predictions (which can describe observation data without primary sources \cite{antip}). One notes that positron datapoints are taken starting from energy 30 GeV. 
At energies below this cut for positrons as well as when step functions are switched on for gamma-ray and antiproton datapoints in \eqref{chispes}, predicted flux is well below observational one. It implies a necessity of different (possible astrophysical) background analysis to describe all the data (at all energies) with the use of our model aimed to explain only PA.

\begin{figure}[t]
	\begin{minipage}{17pc}
		\includegraphics[width=17pc]{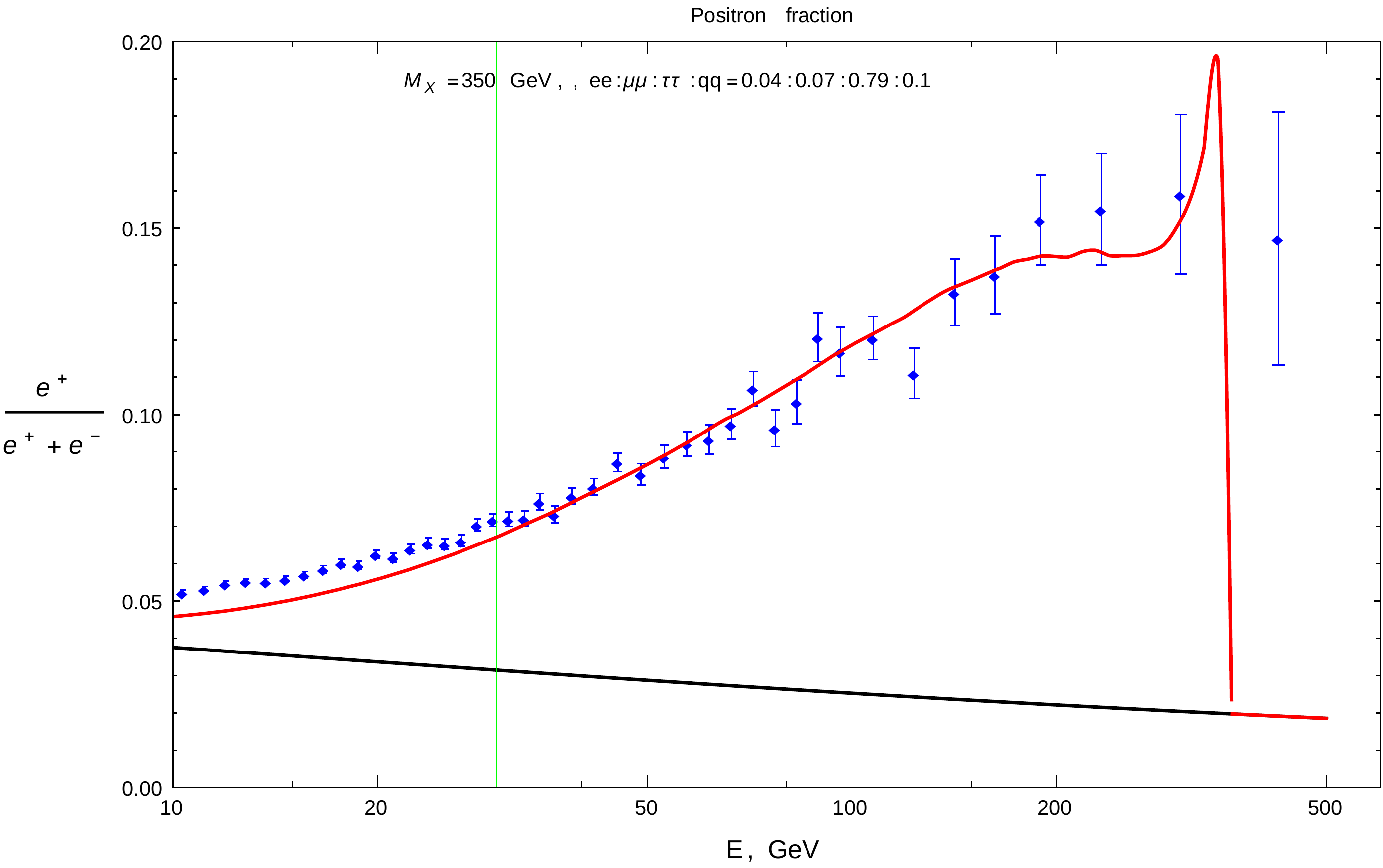}
	\end{minipage}\hspace{2pc}%
	\begin{minipage}{17pc}
		\includegraphics[width=17pc]{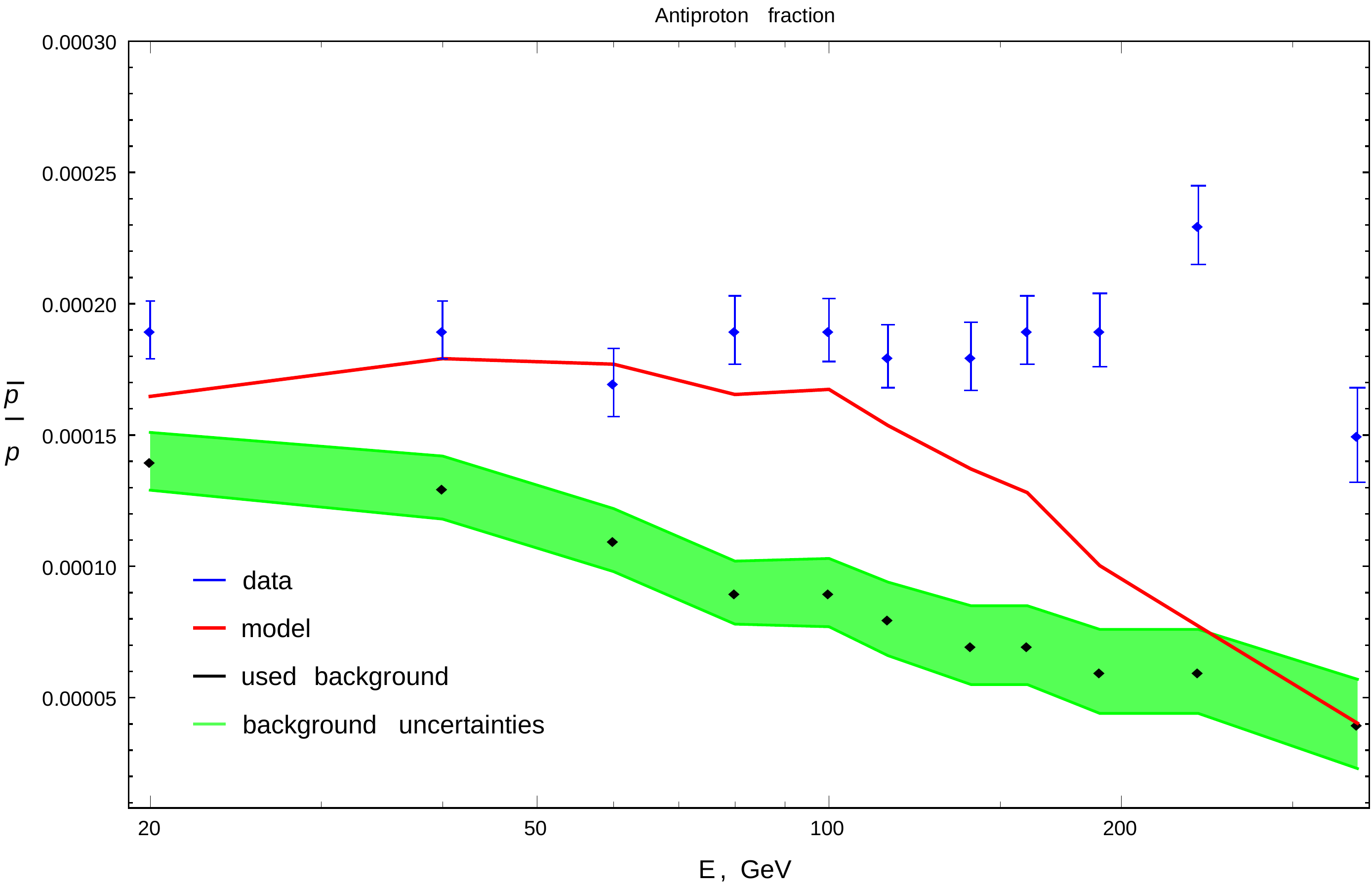}
	\end{minipage}
	\begin{minipage}{17pc}
		\includegraphics[width=17pc]{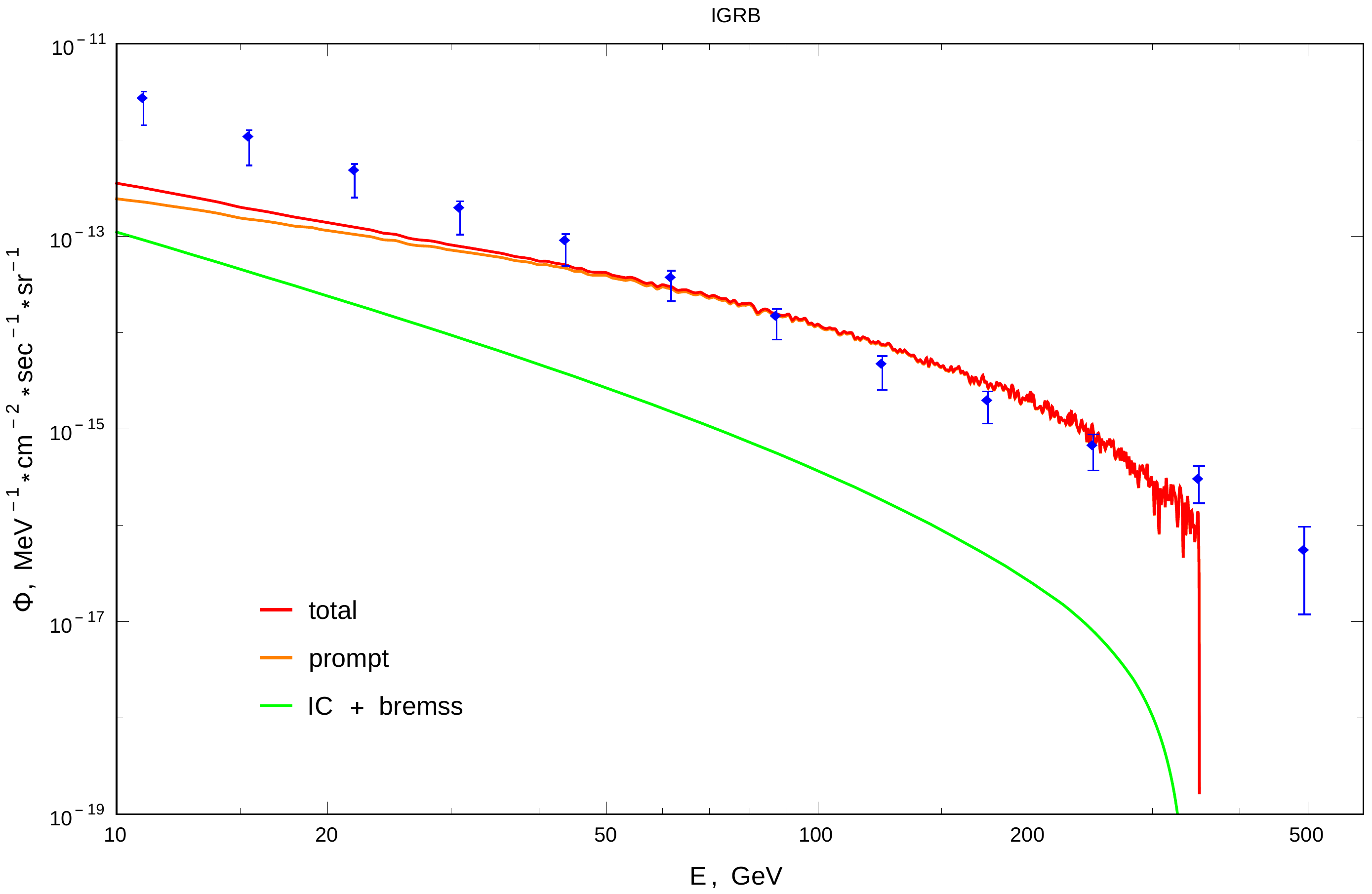}
	\end{minipage}\hspace{2pc}%
	\begin{minipage}{17pc}
		\includegraphics[width=17pc]{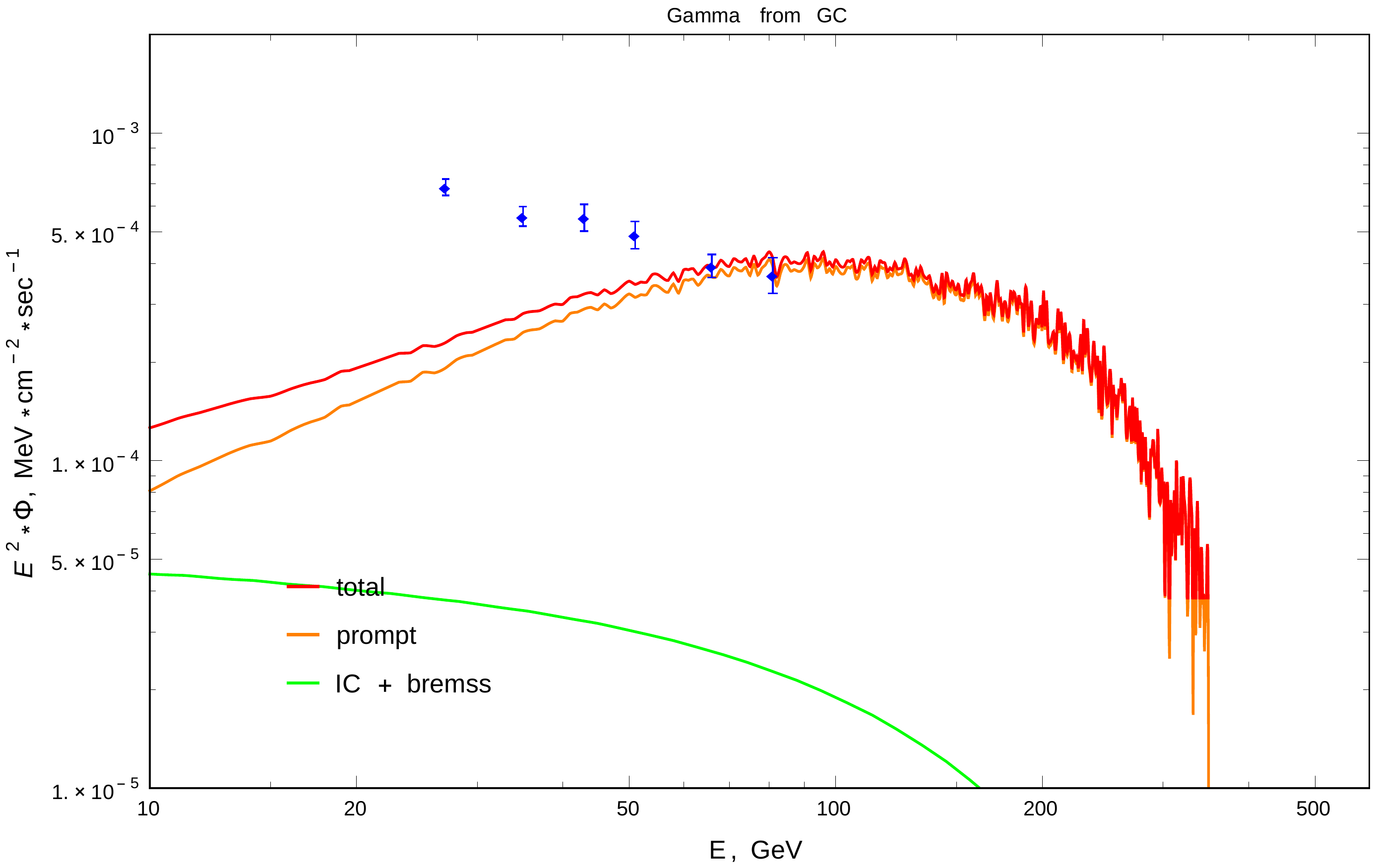}
	\end{minipage}
	\caption{Best fit case for 2 body mode ($m=350$ GeV): positron (top left) and antiproton (top right) fractions, IGRB (bottom left) and gamma from GC (bottom right). Branching ratios are pointed out in the first plot.}
	\label{350}
\end{figure}

\begin{figure}[t]
	\begin{minipage}{17pc}
		\includegraphics[width=17pc]{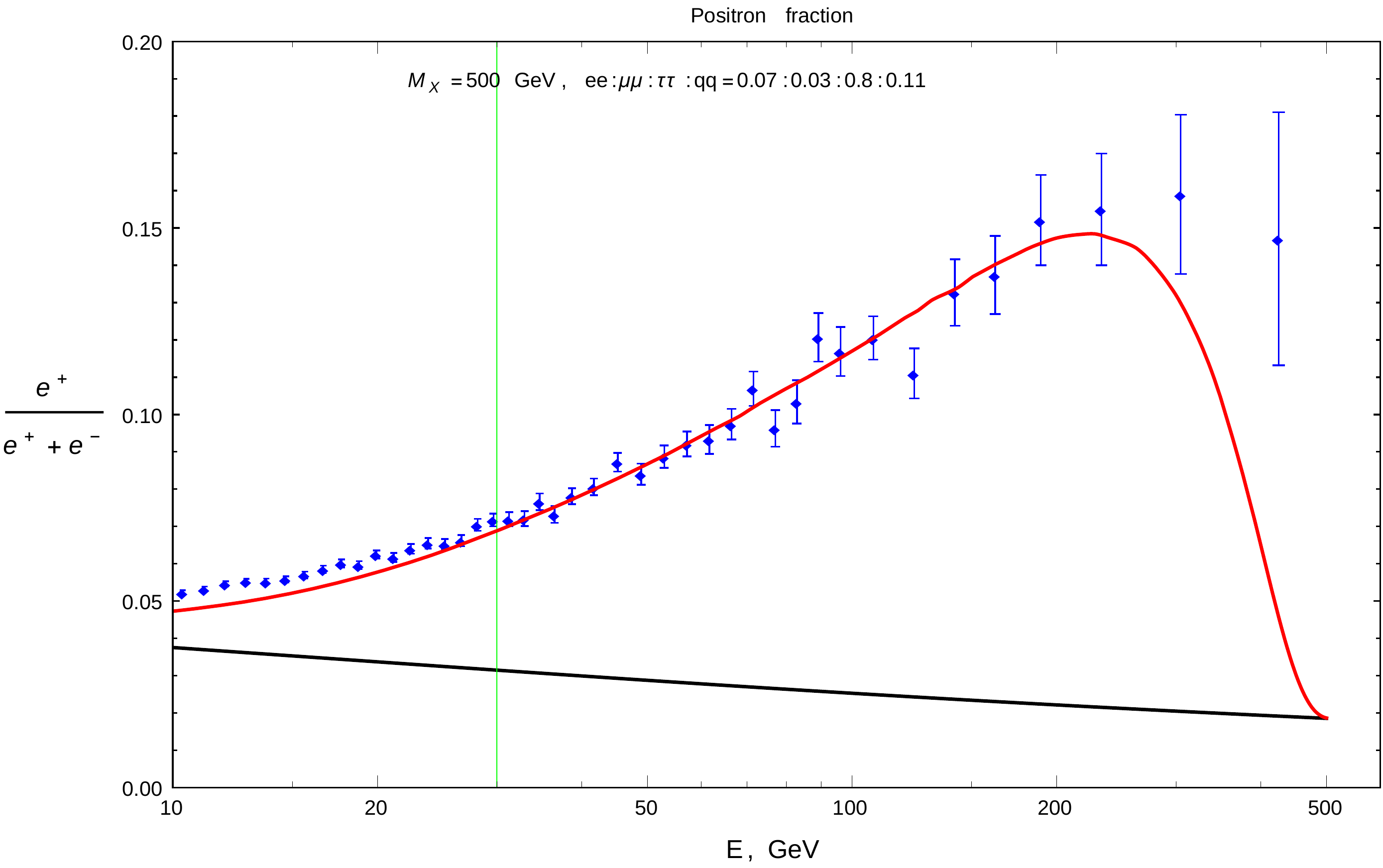}
	\end{minipage}\hspace{2pc}%
	\begin{minipage}{17pc}
		\includegraphics[width=17pc]{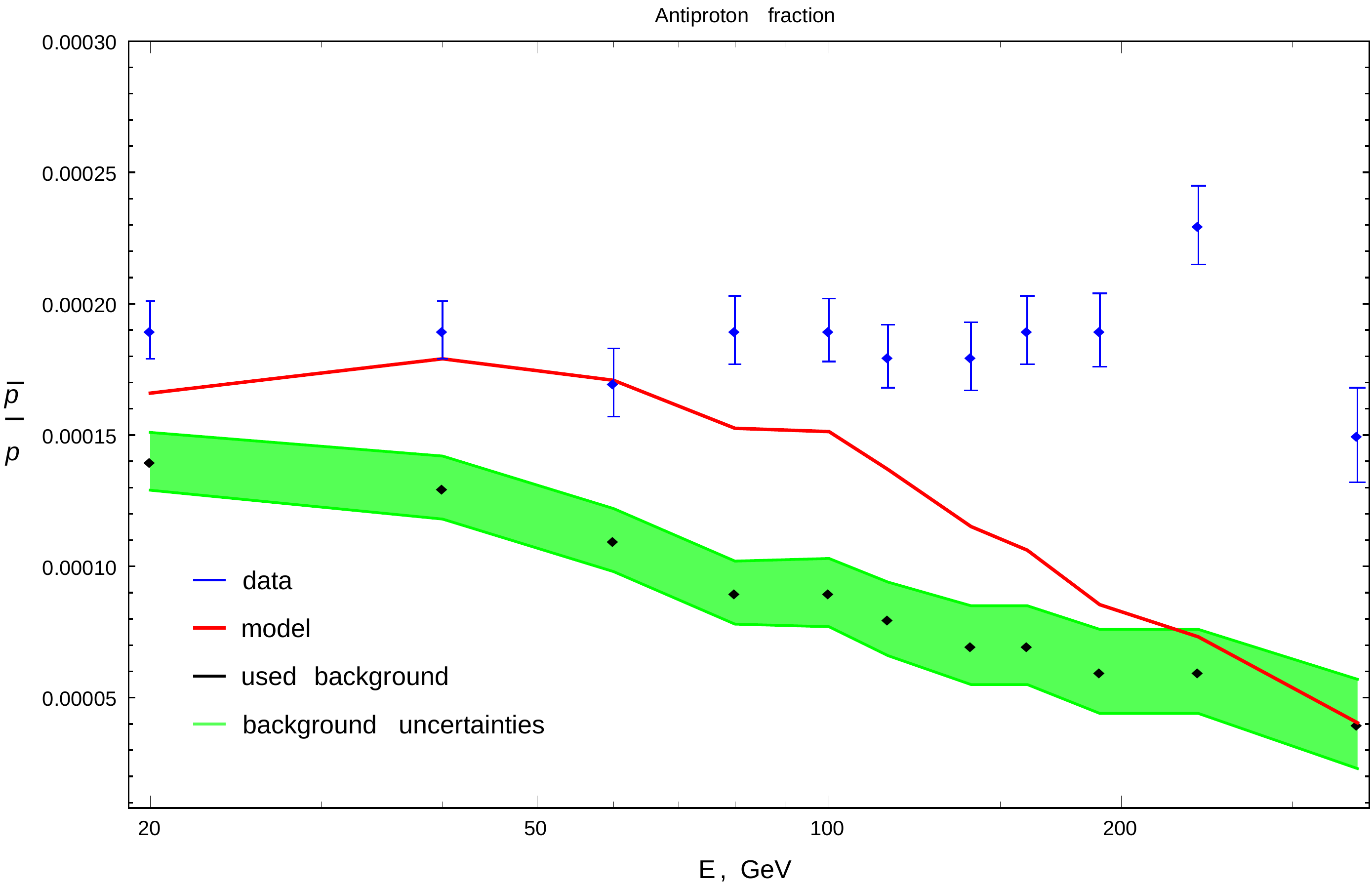}
	\end{minipage}
	\begin{minipage}{17pc}
		\includegraphics[width=17pc]{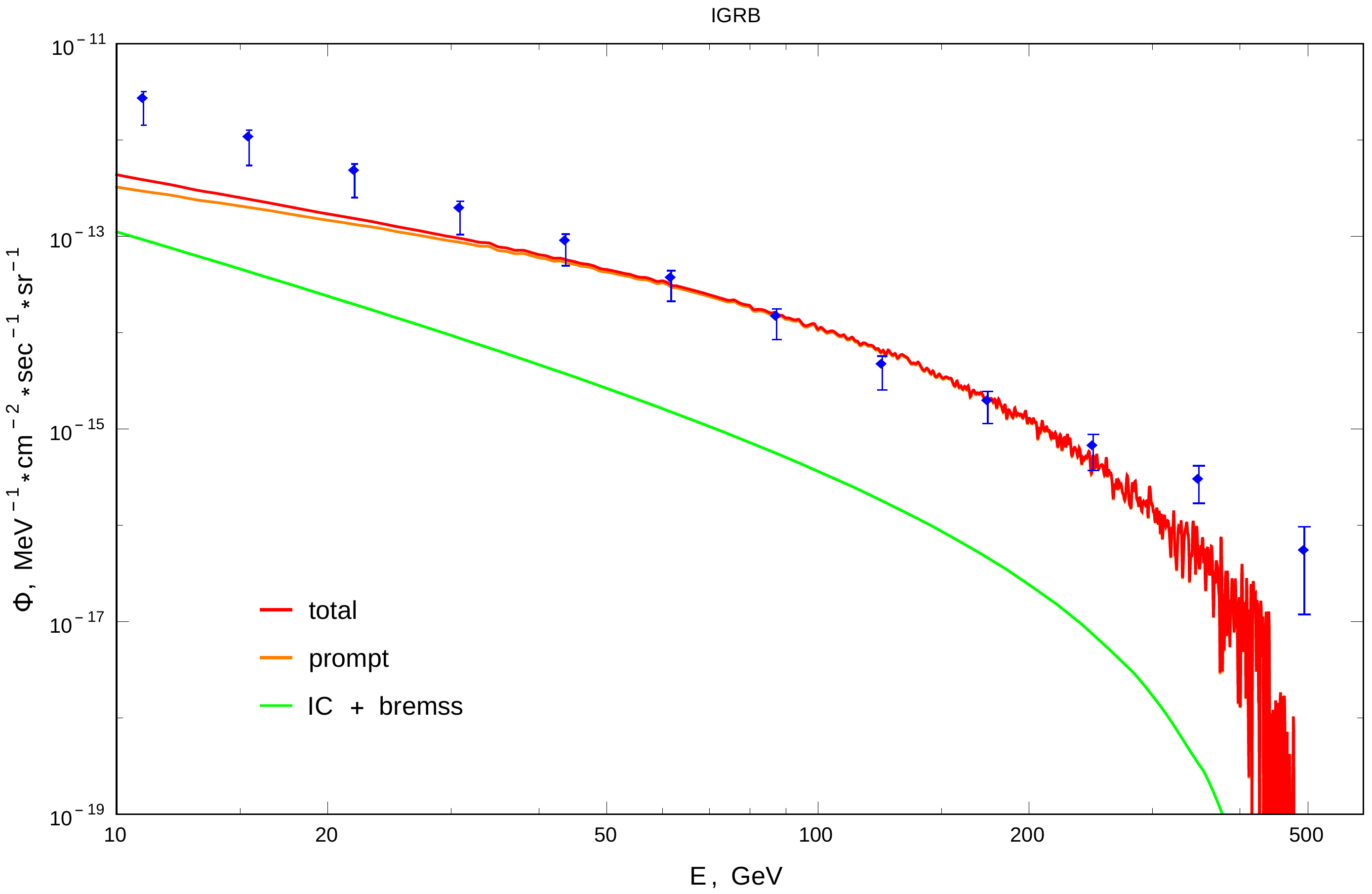}
	\end{minipage}\hspace{2pc}%
	\begin{minipage}{17pc}
		\includegraphics[width=17pc]{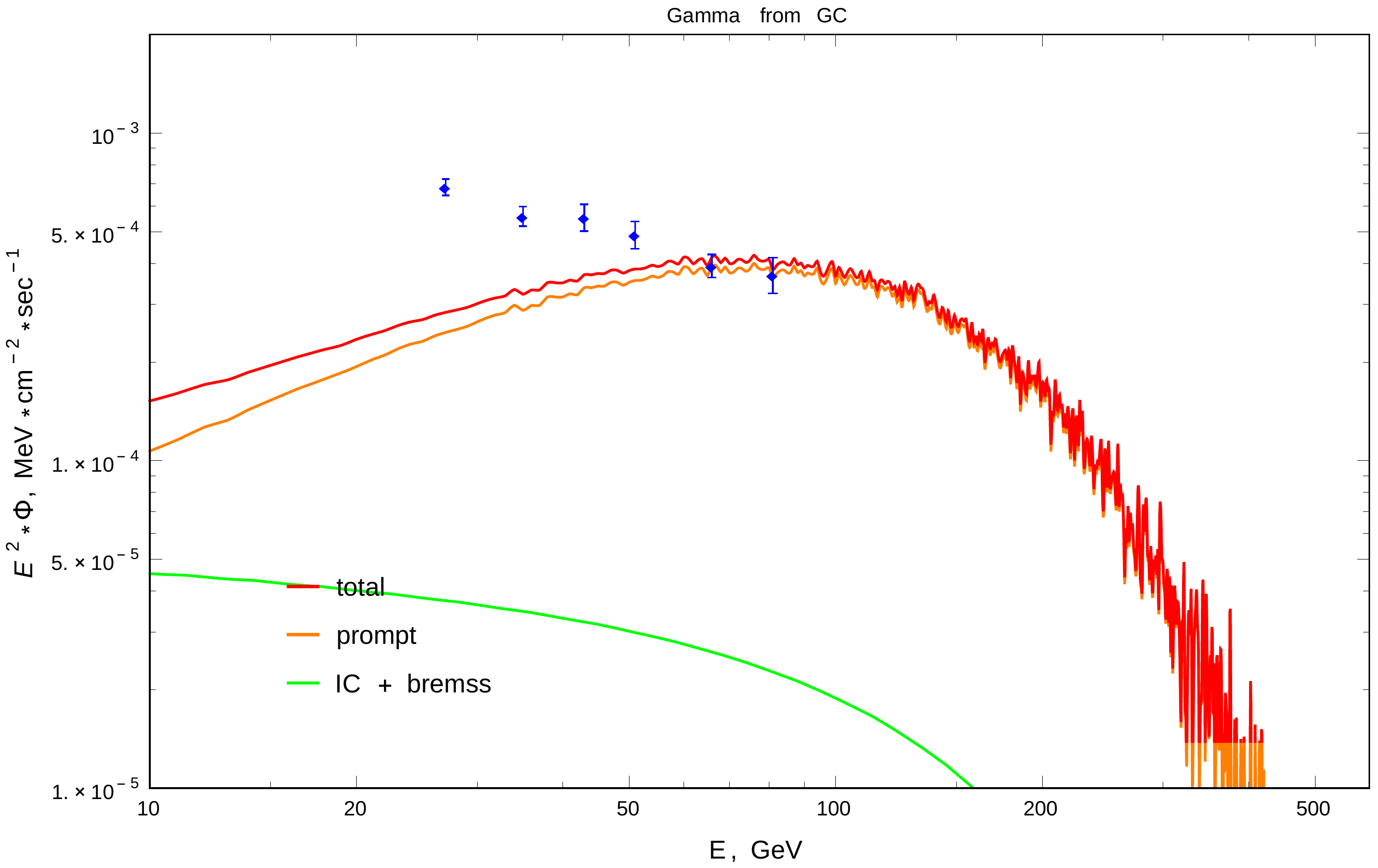}
	\end{minipage}
	\caption{The same as in figure \ref{350} but for 4 body mode ($m=500$ GeV).}
	\label{500}
\end{figure}
    
Branching ratios ($Br_e$, $Br_{\mu}$, $Br_{\tau}$, $Br_q=1-Br_e-Br_{\mu}-Br_{\tau}$) and local emissivity $j=\frac{1}{4}n_\text{loc}^2\langle\sigma v\rangle$ are the varying parameters. The mass of DM particle in case of 2 body mode was taken to be $m=350$ GeV, what corresponds to our best value in previous analysis, and $m=350$, $500$, $750$, $1000$ and $1500$ GeV for 4 body mode. Four body mode was simulated in Pythia through the process $H\rightarrow ZZ \rightarrow f\bar f f\bar f$, where $m_H=2 m$ and $m_Z\gg m_H$ were set. 
Quark mode was taken in form of mixture $2q=(u\bar u+d\bar d)/2$ and $4q=(u\bar u u \bar u+d\bar d d\bar d)/2$ for 2- and 4-body modes respectively.
 
Mass 500 GeV was found to be the best one for 4-body mode, 
%
%
%
%
%
and both 2- and 4-body modes fit data better with inclusion of respective quark modes. Figures \ref{350} and \ref{500} show CR spectra for positron fraction, IGRB, gamma from GC and antiprotons, as expected at the best fit parameters (with quark mode being switched on), in comparison with respective observation data. Gamma-radiation induced by DM has two contributions: as direct products of annihilation (prompt) and as a result of $e^{\pm}$ interaction with galactic medium (inverse Compton scattering, bremsstrahlung, etc.), calculated by GALPROP.

Table \ref{table} shows $\chi^2$ values for the cases with and without quark modes with the mentioned above masses. The number of degrees of freedom $N_{dof}$ by which $\chi^2$ is divided includes $N_\text{tot}=N_\text{pos}+N_\text{IGRB}+N_\text{GC}$ if quark mode is off and antiprotons are not considered, and if they are on $N_{dof}$ includes both $N_\text{tot}+N_\text{antip}$ and only $N_\text{tot}$, for which $\chi^2/N_{dof}$ is written in the Table in brackets to see explicitly the change of quality of data description (positron fraction, IGRB and gamma from GC) with inclusion of the quark mode. Note that all the pure (single final species) modes give basically worse $\chi^2$ values.

\begin{table}[t]
	\begin{tabular}{|c|c|c|c|c|}
		\hline
		&$2e+2\mu+2\tau$&$2e+2\mu+2\tau+2q$&$4e+4\mu+4\tau$&$4e+4\mu+4\tau+4q$\\
		\hline
		$\chi^2/N_{dof}$ & 1.75 & 1.28  (1.67)& 1.14 & 0.789 (1.03) \\
		\hline
	\end{tabular}
	\caption{Values of $\chi^2$ for different annihilation channels in the best fit cases. The value in the brackets does not take into account number of antiproton datapoints in $N_{dof}$.}
	\label{table}
\end{table}

\begin{figure}[p]
	\begin{minipage}{17pc}
		\includegraphics[width=17pc]{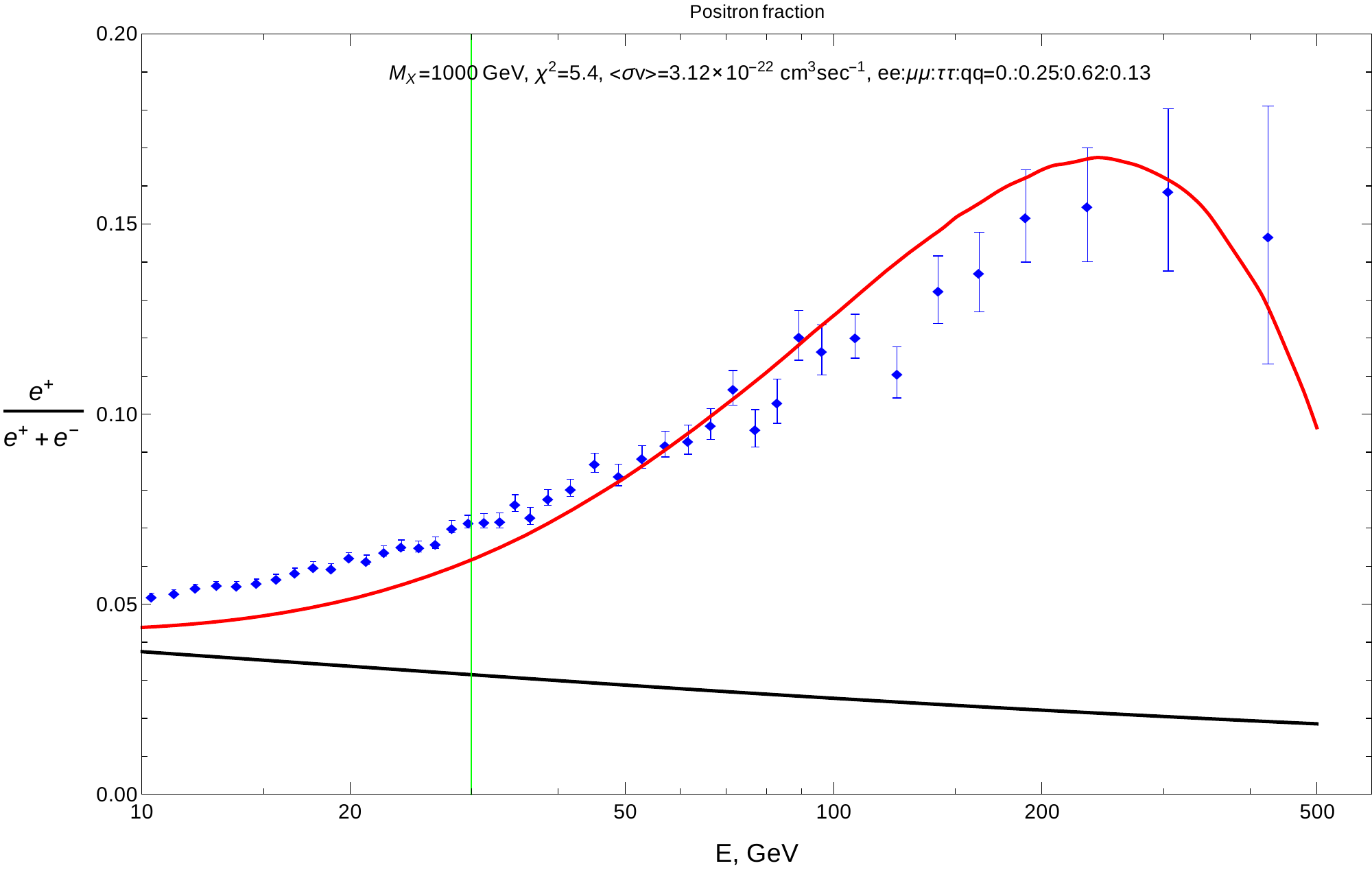}
	\end{minipage}\hspace{2pc}%
	\begin{minipage}{17pc}
		\includegraphics[width=17pc]{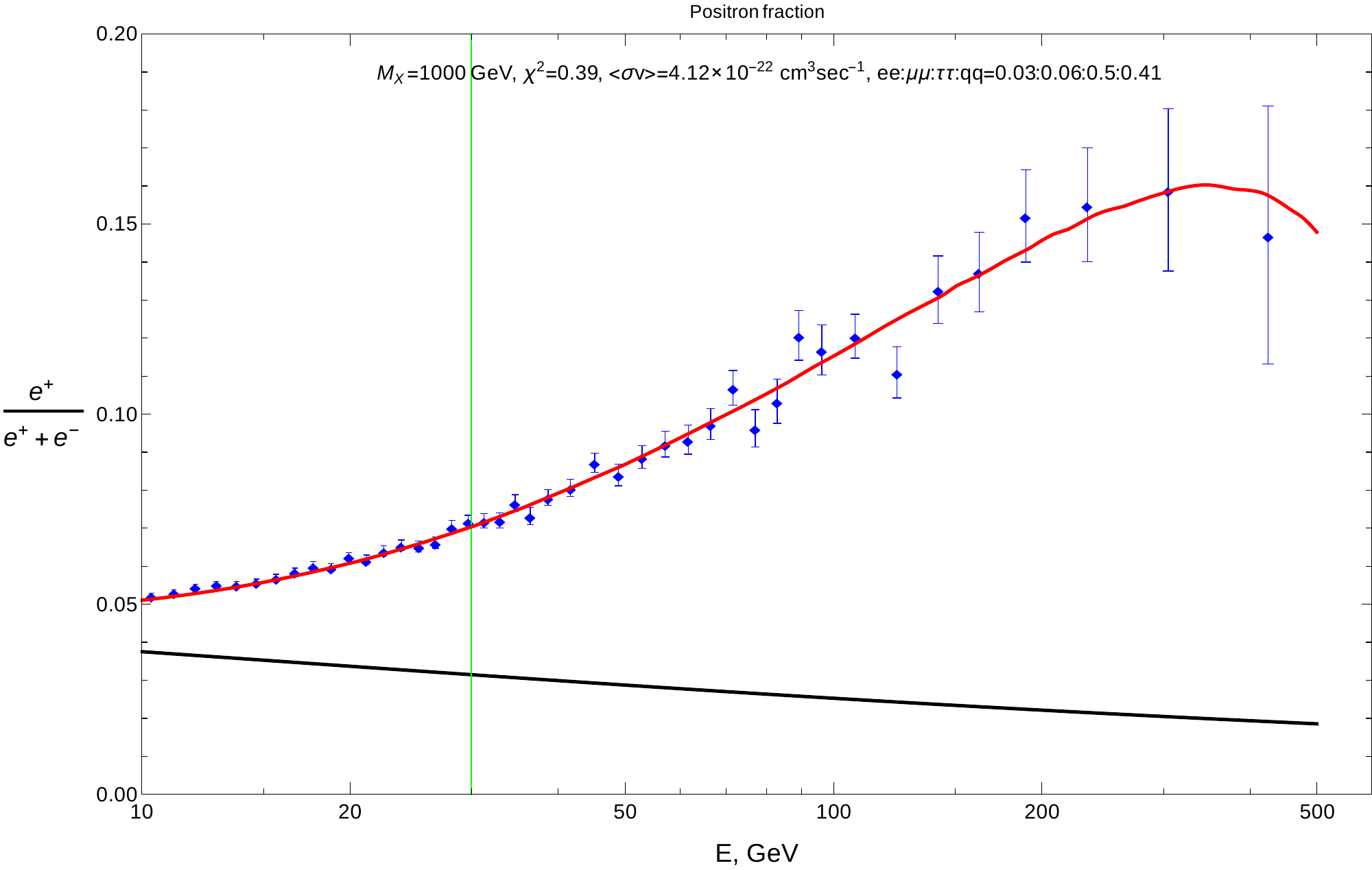}
	\end{minipage}
	\begin{minipage}{17pc}
		\includegraphics[width=17pc]{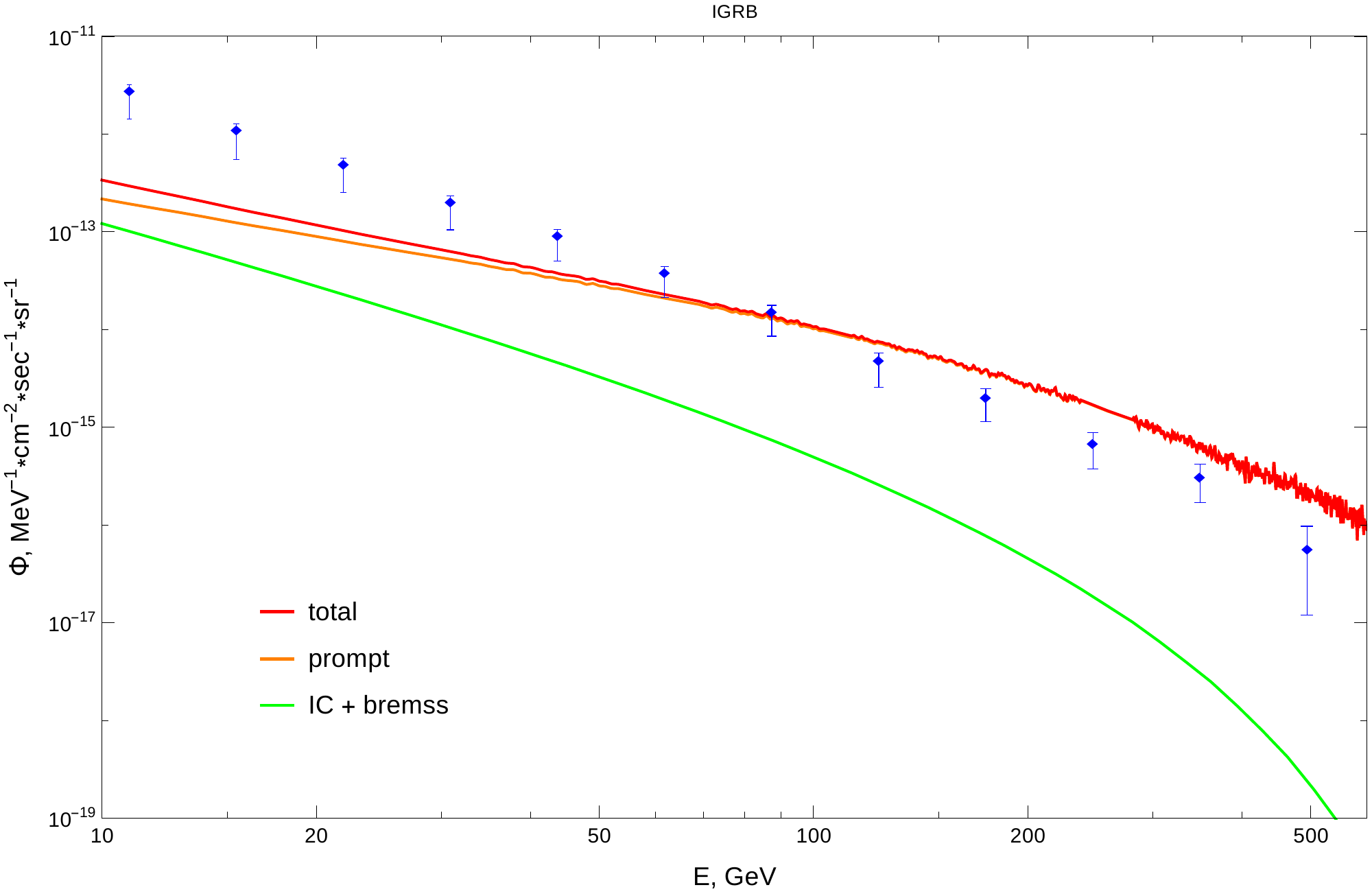}
	\end{minipage}\hspace{2pc}%
	\begin{minipage}{17pc}
		\includegraphics[width=17pc]{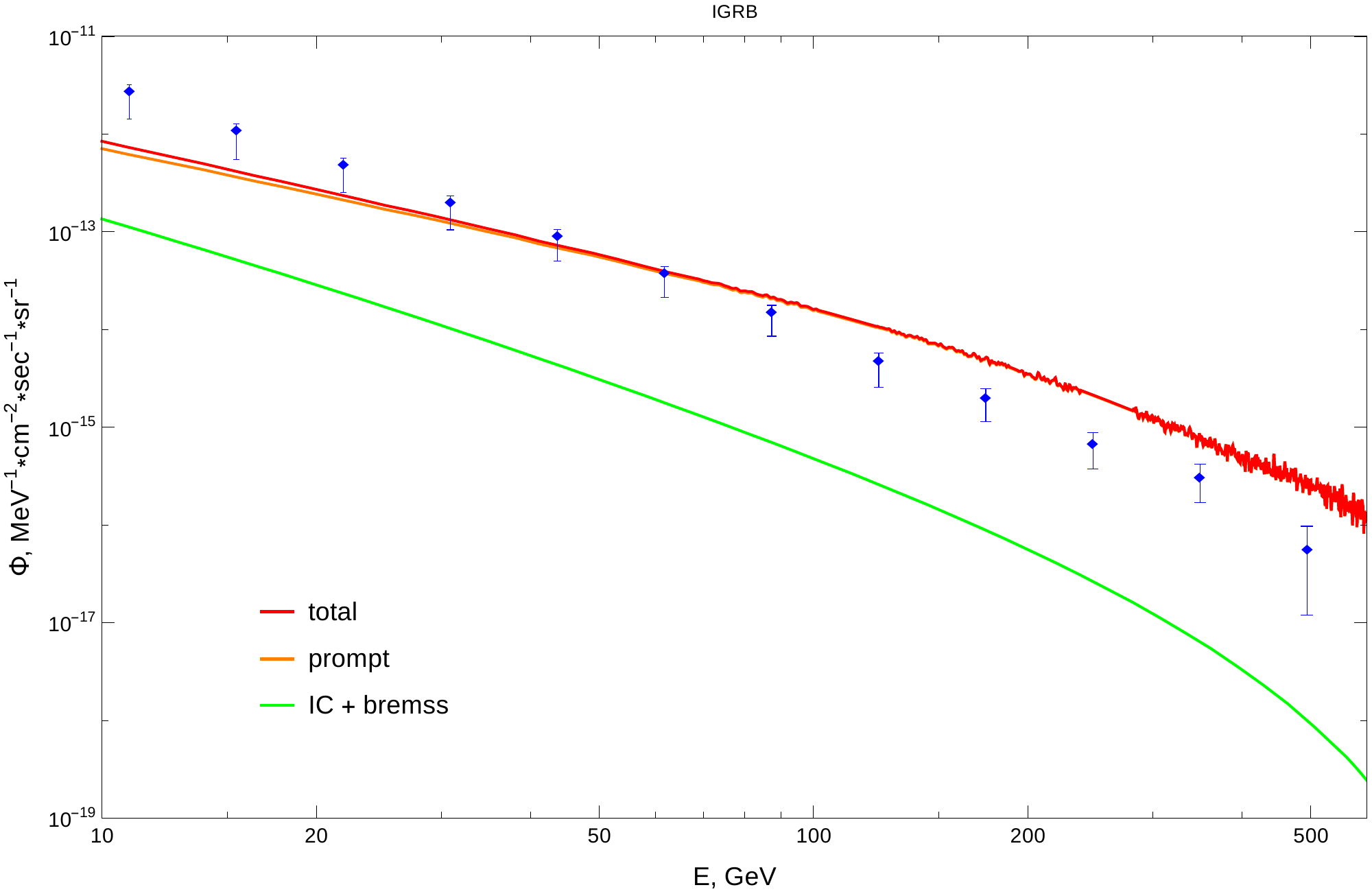}
	\end{minipage}
	\begin{minipage}{17pc}
		\includegraphics[width=17pc]{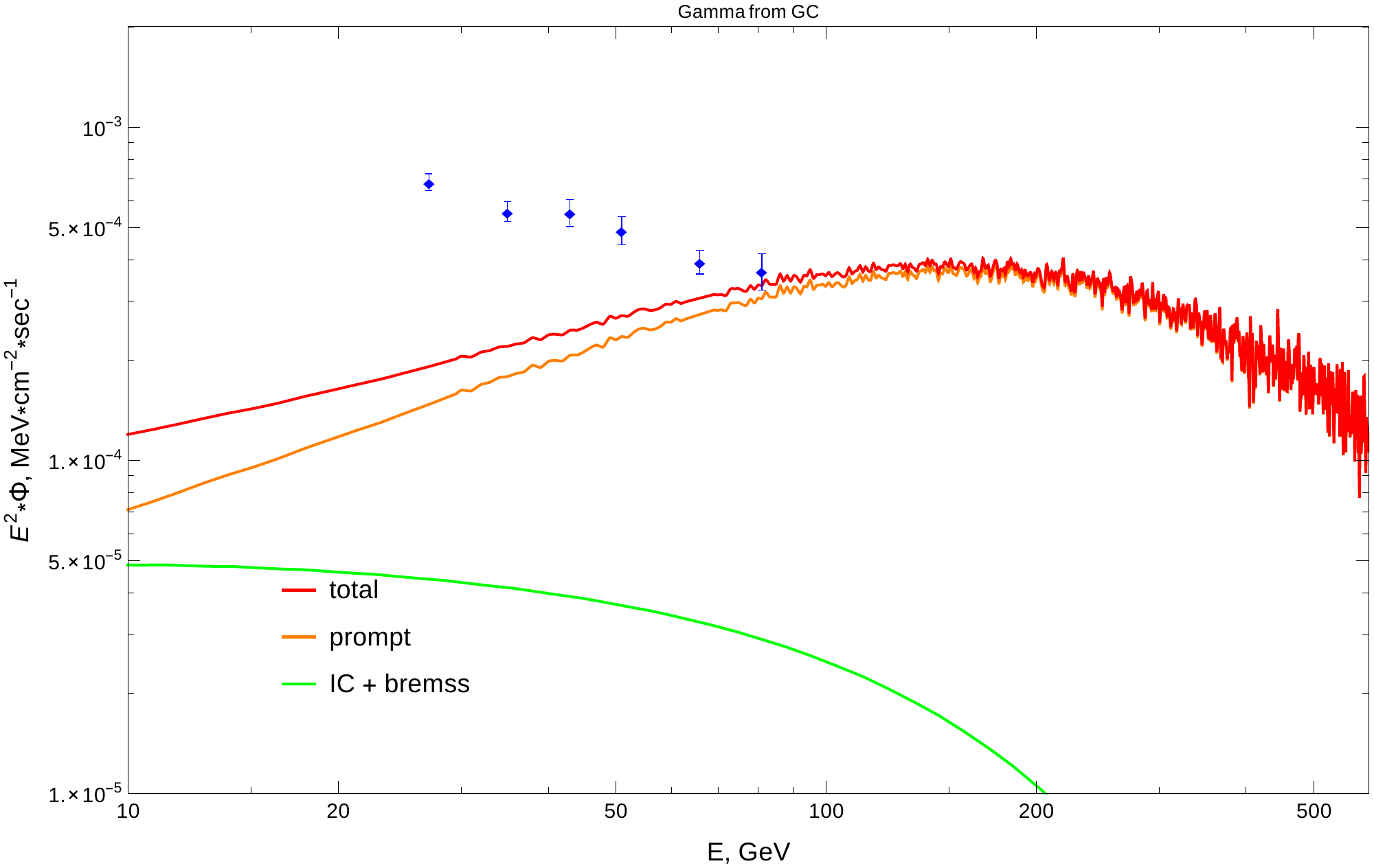}
	\end{minipage}\hspace{2pc}%
	\begin{minipage}{17pc}
		\includegraphics[width=17pc]{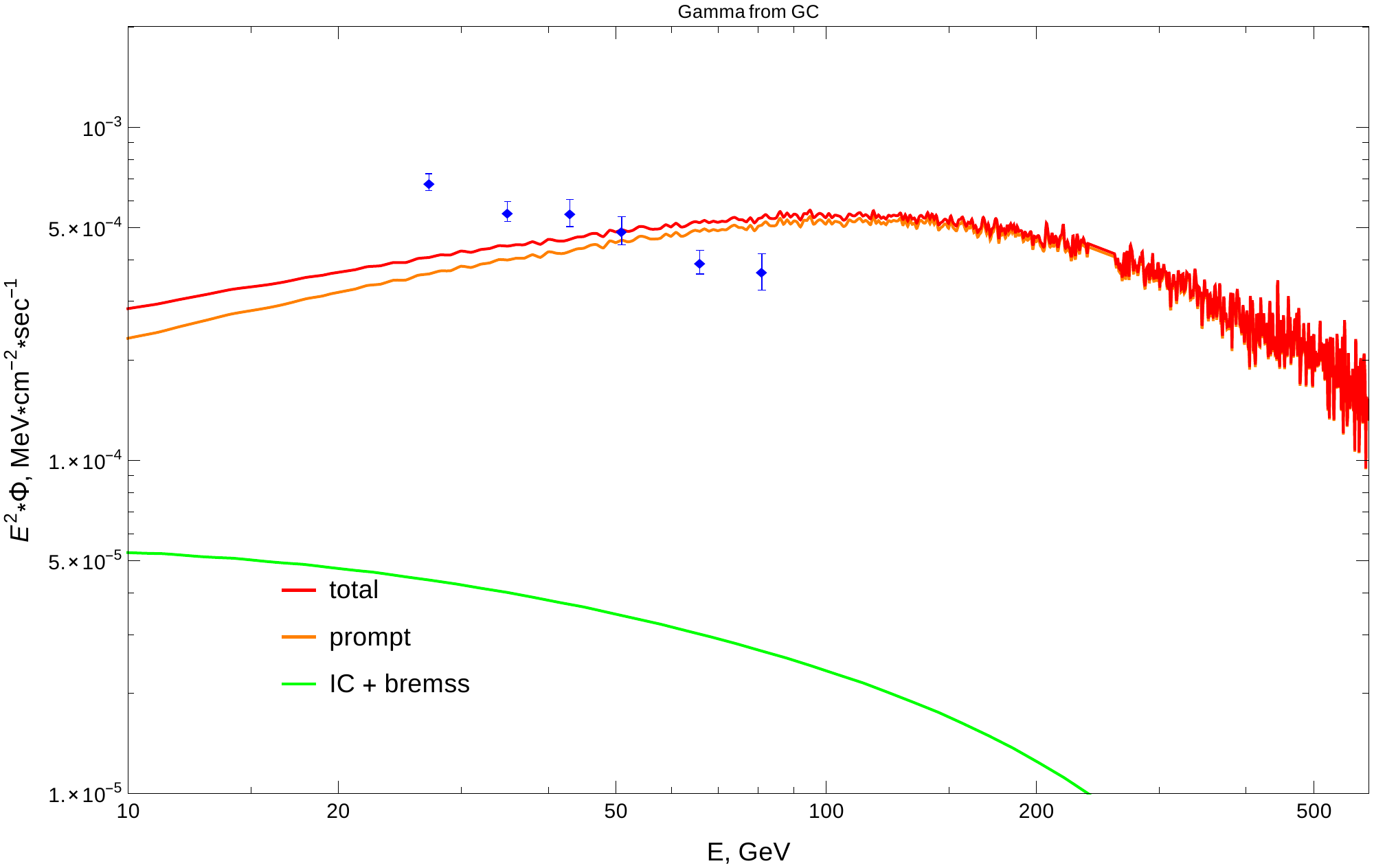}
	\end{minipage}
	\begin{minipage}{17pc}
		\includegraphics[width=17pc]{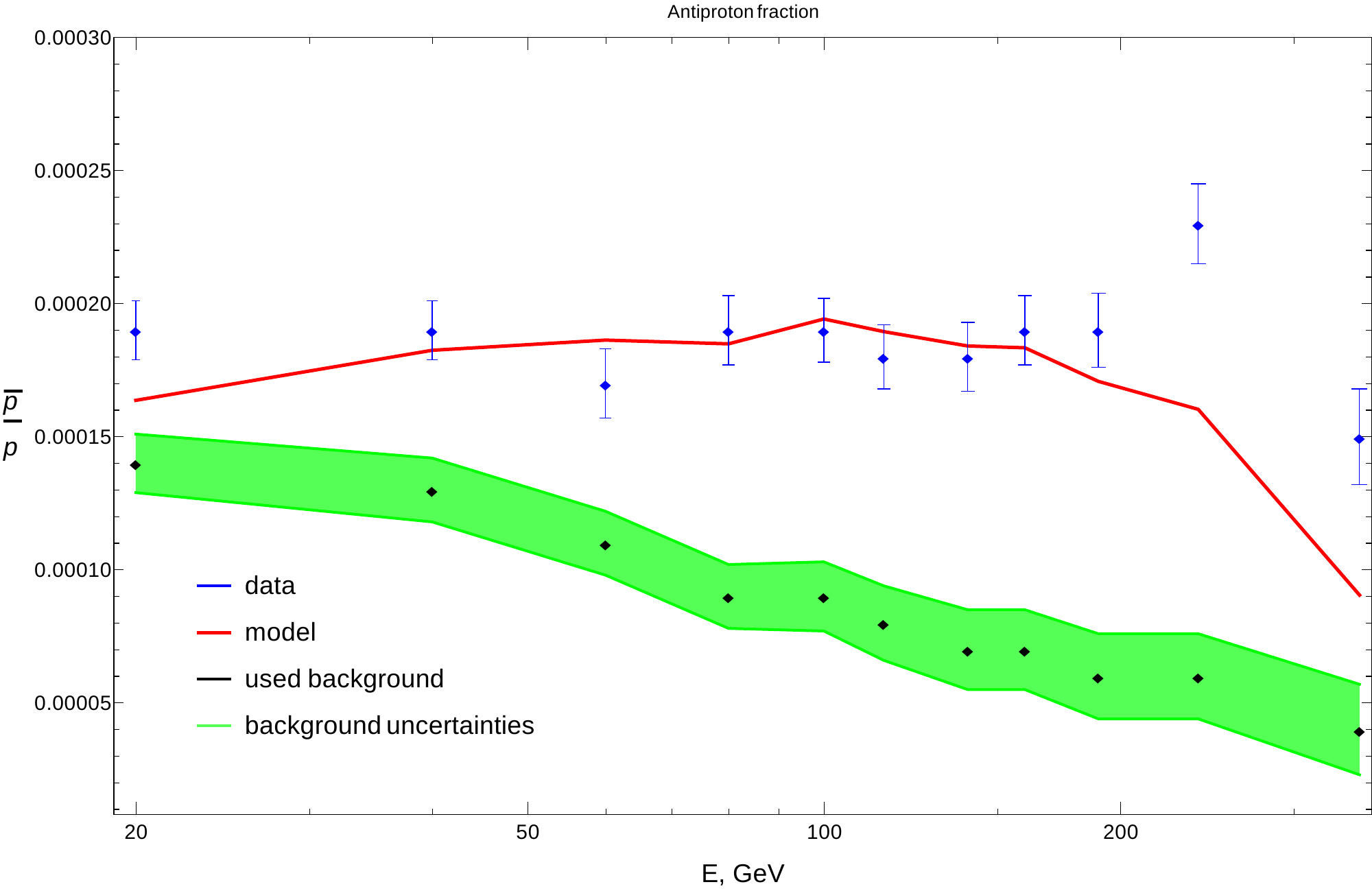}
	\end{minipage}\hspace{2pc}%
	\begin{minipage}{17pc}
		\includegraphics[width=17pc]{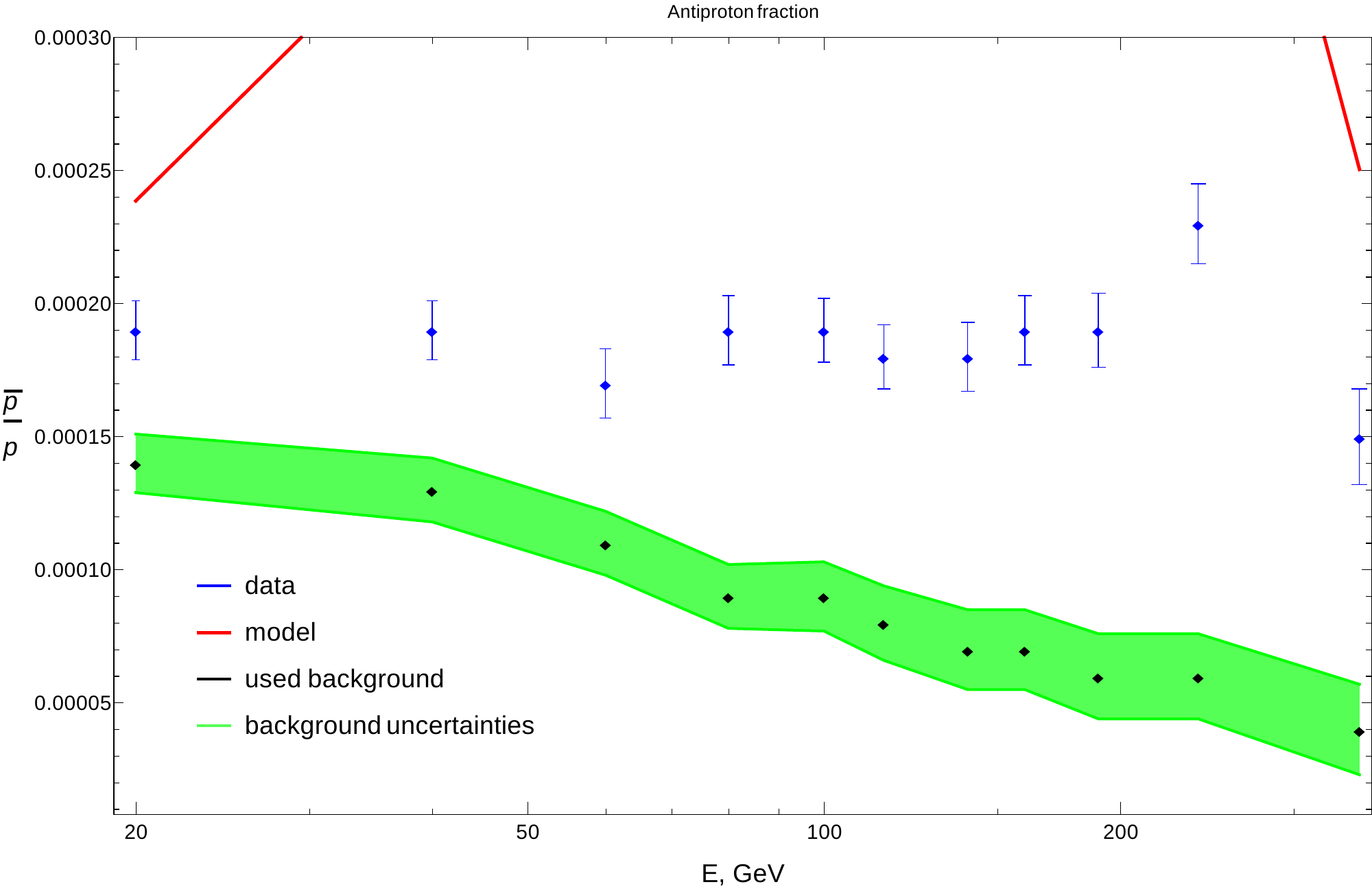}
	\end{minipage}	
	\caption{The case for $m=1000$ GeV and 4-body mode obtained by fitting all data (left) and only data on positrons (right). From top to bottom: positrons, IGRB, gamma from GC, antiprotons.}
	\label{1000}
\end{figure}

It is worth to note that accuracy of AMS-2 positron datapoints is high at middle and low energies. Fitting them regulates (increases) branching ratios of $\tau$- and quark-modes, which give relatively soft positrons. But from other side, these modes produce many gammas, that is why even in disk case, data on gamma-radiation restricts the model. However in case of dark halo, descripancy with observation of gamma-radiation arises even for $e$- and $\mu$-modes \cite{dd-1, ML-1, ML-2}. Ignorance of gamma would allow, due to bigger branching ratios of $\tau$- or/and quark-modes and also higher DM particle mass, to describe the measured positron fraction considerably better.
To illustrate influence of gamma (and antiprotons), we show in Fig. \ref{1000} the best fit positron fractions obtained for $m=1000$ GeV taken into account both data on gamma-radiation and antiprotons (left plots) and not, fitting only positrons (right plots). In the first case $\chi^2/N_{dof}=5.4$ if $N_{dof}$ includes $N_\text{tot}+N_\text{antip}$ and $10.2$ if it includes only $N_\text{pos}$, in the second case $\chi^2/N_{dof}=0.73$ ($N_{dof}$ includes only $N_\text{pos}$) and it contradicts strongly to data on gamma and antiprotons (the latter could be easily eliminated by switching off quark mode). It can be seen in Fig. \ref{1000}. Note, that in the first case antiprotons are described much better, while the global $\chi^2$ is worse.

Recently there appeared results of Fermi-LAT \cite{bib:FermiLAT1, bib:FermiLAT2, bib:FermiLAT3} that unresolved point-like extragalactic sources (basically blazars) could explain essential part of IGRB \cite{fermi-igrb}. We discussed it earlier \cite{dd-1}. Now one could add that Fermi-LAT results relate to integrated gamma-ray flux for $E>50$~GeV ($>10$~GeV in the last aforementioned work of Fermi), while the main DM contribution to IGRB is around $E_{\rm DM}\sim 100-200$~GeV. 
At decreasing power-law of the flux, $\propto E^{-2.5}$, total flux at energies $E>E_{\rm DM}$~GeV is as small as $0.35\cdot(100 \text{ GeV}/E_{\rm DM})^{1.5}\sim 0.12-0.35$ of total flux at $E>50$~GeV, what is consistent with Fermi results (accordingly to which it must be $<0.28$ at $1 \sigma$). So the flux at energies from $50$~GeV to $E_{\rm DM}$ is to be dominantly explained by unresolved sources. In so manner, accounting for them could not only stronger constrain DM contribution but improve description of IGRB as a whole.
    
    \section{Conclusion}
In these proceedings we continued our analysis of possibility to explain positron anomaly with dark matter. Ordinary scenario conflicts with observation of gamma-radiation. To avoid it we consider a dark disk model. Here we opened 4-body and quark channels of DM annihilation and extended observation data base due to cosmic antiprotons. All the observational data (on cosmic positrons, IGRB, gamma from GC and antiprotons) are simultaneously fitted. It provides the most flexible possibility to test model.

In the result we obtained that accounting for data on antiprotons gives no model restrictions having switched on quark mode. Attempts to fit a possible antiproton excess may be promising, however higher DM particle mass is most likely required for it what will restore strong tension with gamma. Though such attempts should involve the play with background (secondary) antiprotons.
Generally 4-body modes are a little better than 2-body ones from viewpoint of data description. Inclusion of quark mode improves both mode cases (otherwise, procedure of minimizing $\chi^2$ \eqref{chiSq} would nullify $Br_q$). The most favourable values of the DM particle mass are 350 GeV and 500 GeV in cases of 2- and 4-body modes respectively.

The obtained best fit values of the branching ratios are not tied to any underlying physical model, our aim here was firstly to probe opportunity of CR puzzle solution in principle.

We restricted ourselves by consideration of only annihilation. Decay case should not differ noticeably.

\section*{Acknowledgements}
The authors express their special thanks to R.~I.~Budaev and M.~N.~Laletin, on the basis of whose work this one is done. Also we would like to thank S.~R.~Rubin for interest to the work and useful discussions.

This work was supported by the Ministry of Education and Science of the Russian Federation, MEPhI Academic Excellence Project (contract \textnumero~02.a03.21.0005, 27.08.2013). The work of K.~M.~B. is also funded by the Ministry of Education and Science of the Russia, Project \textnumero~3.6760.2017/BY. The work of A.~A.~K. was supported by Russian Science Foundation (\textnumero~15-12-10039).


\end{document}